# Random packing fraction of binary hyperspheres with small or large size difference: a geometric approach

*H.J.H. Brouwers* ,
*Department of the Built Environment, Eindhoven University of Technology*
*P.O. Box 513, 5600 MB Eindhoven, The Netherlands*
*Contact address: jos.brouwers@tue.nl*
(Dated: August 8, 2025)

The random packing fraction of binary particles in D-dimensional Euclidean space $\mathbb{R}^D$ is studied using a geometric approach. First, the binary packing fraction of assemblies with small size difference are studied, using a recently developed model that has its foundations in the excluded volume model by Onsager for cylinders and spherocylinders (D = 3). According to this model the packing increase by bidispersity is proportional to $(1 - f)(u^D - 1)^2$, with f as monosized packing fraction, u as size ratio and D as space dimension. The model predictions are compared with computational results for disks in two dimensions (D = 2) and hyperspheres in the large-dimension limit (D → ∞), yielding very good agreement. Subsequently, the packing of hyperspheres with large size difference is modeled, employing the classic theory of Furnas. This theory, developed for three dimensions, starts from an infinite size ratio of larger and smaller particles (u → ∞). Here, the pertaining equations are applied to hyperspheres, and successfully compared with computational results for hyperspheres in the large-dimension limit. Furthermore, an asymptotic approximation of the binary packing fraction for large size ratio is derived, which shows that the first order variation of the Furnas packing fraction ($u^{-1}$ = 0) is proportional to $(2 - f)u^{-1}$. Finally, a normalized D-dimensional binary packing graph is presented, governing a simplified phase diagram that borders the binary random packing fraction of amorphous assemblies. To summarize, basic space-filling and geometric ("athermal") theories on "simple" hard spheres appear to be a valuable tool for the study of hyperspheres' random packing and amorphization.

## 1. INTRODUCTION

The packing of particles is an old physical and mathematical puzzle and has received much attention the past millennia [1]. Attention has for instance been paid to revealing packing geometries and the route to understanding liquids and (amorphous) materials. Hard sphere systems are ideal to study liquid-glass-crystal transitions [2]. Furthermore, study of amorphous hyperspheres in D-dimensional space enables a better understanding of glass formation in three dimensions, and it brings the problem in contact with signal digitization and error coding theory [3].

When equally shaped particles with different sizes are randomly packed, *i.e.* generating a polydisperse packing of similar particles, the packing fraction increases compared to the monosized packing of the congruent (or identical) particles. By combining two similar particles of different sizes, such a polydisperse packing can readily be assembled. In this paper this specific polydisperse particle packing is analysed, *viz.* the packing of two discretely sized and equally shaped particles, here termed bidisperse or binary mixtures. Though this bidisperse packing is a relatively simple polydisperse system, it forms the basis of the packing description of polydisperse arrangements. Early work on binary packings was for instance aimed at constructing packings of continuously sized particles with a wide size ratio [4-9].

The binary packing of similar particles was studied experimentally, computationally and analytically [10-32]. For binary mixes with size ratio u close to unity (u ↓ 1), analytical equations are available [18, 20, 28, 32]. Also, for the other limit, *viz.* infinite size ratio u (u → ∞), *i.e.* two noninteracting fractions, an analytical expression for the binary void fraction is available [4, 5], revisited later [22, 25].

Here, in Section 2 first the model for binary particles with small size disparity is introduced [32], which was based on Onsager's excluded volume model from 1949 [33]. Onsager developed this original geometric model for the isotropic liquid-to-nematic (I-N) phase transition of hard rodlike (spherocylinders and cylinders) particles, which was published in his seminal paper. Onsager demonstrated that a phase transition can be predicted based on two-particle (spherocylinders or cylinders) interactions represented by the second virial term in an expansion of the free energy of the system. Onsager based these expressions on the orientally averaged excluded volume of two spherocylinders or two cylinders with unequal lengths and diameters. In essence this is an example of a statistical geometric approach. In [32], this excluded volume concept of two-particle pairs was combined with the statistically probable combinations of small and large particle pairs, yielding an analytical expression for the packing fraction of binary similar particles with small size disparity. This geometric approach of particle packing was successfully validated against a broad collection of computational and experimental data of packings in three dimensions. Here, this model is applied to binary disks in plane and to binary hyperspheres with small size difference ($u^D$ ↓ 1).

Next, in Section 3, the classic model of Furnas is recapitulated, which formulates the packing of particles with infinite size difference (or $u^{-1} = 0$). About 100 years ago, Furnas [4, 5] introduced the concept of noninteracting particle classes, *i.e.* particle groups where the smallest particle of one group is much larger than the largest particle of the other group. Combining the groups implies that they are not interacting and forming separate phases. This geometric concept has been proven to be correct by experiments [4, 5, 9, 11] and simulations [22, 25, 27] in three dimensions. Here, we will apply this model to binary mixes of hyperspheres with infinite large size ratio. In Section 4, an asymptotic approximation of the bidisperse fraction is presented for large but finite size differences (so $u^{-1} > 0$), that approaches the Furnas solution in the infinite size difference limit.

Subsequently, in Section 5 a generic graph is introduced of the normalized bidisperse packing fraction. This figure borders the normalized packing fraction of amorphous assemblies, as function of composition and from size ratios unity to infinity, and in the vicinity of these -opposite- limits. The conclusions are collected in Section 6.

The presented models provide the random or amorphous packing of nonoverlapping (*i.e.* hard) particles in the bidisperse case relative to the monodisperse case, in D-dimensional Euclidean space $\mathbb{R}^D$. They are applicable when the packing of the smaller and the larger particles, and their binary packing, are compacted equally. Whether the assembly's density corresponds to the maximally random jammed state (MRJ) [1], random close packing (RCP), random loose packing (RLP), or a configuration in between these closest or loosest possible ways of particle packing, is inconsequential. Random particle packings are prototypical glasses in that they are maximally disordered while simultaneously being mechanically rigid. Moreover, size dispersity frustrates crystallization and is therefore a glass phase enabler. Indeed the glass transition is related to a specific packing density, in "Table II" [2] packing fraction values for different protocols are listed. Also these packing fractions are affected by bidispersity and are captured by the presented model.

## 2. SMALL SIZE DIFFERENCE

This paper addresses the assemblies of binary (discretely sized) similar particles in D-dimensional space, the larger and smaller ones with characteristic sizes $d_L$ and $d_S$, respectively, with a normalized number distribution

$$P(d) = X_S \, \delta(d - d_S) + X_L \, \delta(d - d_L) \quad , \tag{1}$$

where $\delta$ is the Dirac delta function, and $X_S$ and $X_L$ are the number fractions of the smaller and larger components for which the following identity holds

$$X_S + X_L = 1 \quad . \tag{2}$$

In this section the analytical model for binary mixtures with small size disparity [32] is recapitulated and subsequently applied to binary hyperspheres in the large-dimension limit.

### 2.1 Analytical model

By employing the excluded volume model of Onsager [33], in [32] the following equation was derived for the random packing fraction of similar binary D-dimensional particles, assuming that mixes and two monodisperse assemblies possess same compaction, and a small size difference $u^D$:

$$\eta(u, X_L, D) = \frac{f(X_L(u^D - 1) + 1)}{X_L(u^D - 1) + 1 - X_L(1 - X_L)(1 - f)v(u, D)} \quad , \tag{3}$$

with $\eta(u, X_L, D)$ as binary packing fraction, $f$ as monosized packing fraction, $u$ as size ratio $d_L/d_S$ and as contraction function (Appendix)

$$v(u, D) = \frac{(u^D - 1)^2 (1 - D^{-1})}{2(u^D + 1)(1 - 2^{1-D})} \quad , \tag{4}$$

and D as the space dimension.

The nominator of Eq. (3) reflects the total volume of the particles, and the denominator the total volume of the packing [32]. Eq. (3) reveals that that the effect of bidispersity on packing fraction is governed by the product $X_L(1 - X_L)(1 - f)v(u, D)$, where $X_L(1 - X_L)$ accounts for the composition, $(1 - f)$ for the monosized void fraction (depending on particle type and densification) and $v(u, D)$ for the contraction function (depending on size ratio and dimension). The contraction function followed from applying the Onsager excluded volume model to uneven particle pairs of spherocylinders and cylinders and assessing their statistical occurrence. It appeared that for larger size ratios, an expression for $v(u, D)$ provided by [10] is more accurate. In [32] it was seen that for both RCP and RLP in $\mathbb{R}^3$ the modified model is accurate up to $u = 2$ or so, so a volume ratio $u^D$ of about 8. In [32] it was furthermore postulated that this expression is also applicable to $D \neq 3$, in the Appendix this is further elaborated on.

The large constituent number fraction is related to the large constituent volume fraction $c_L$ by:

$$X_L = \frac{c_L}{(1 - c_L)u^D + c_L} \quad , \tag{5}$$

so that Eq. (3) can be written as

$$\eta(u, c_L, D) =$$

$$\frac{f(c_L(1-u^D)+u^D)}{c_L(1-u^D)+u^D - c_L(1-c_L)(1-f)v(u,D)} \quad . \quad (6)$$

Eqs. (3), (4) and (6) reveal that the increase in packing by binary dispersity is governed by volume ratio $u^D$ (for D = 2 it constitutes the surface area ratio) of a large and a small particle, so the size ratio to the power dimension.

### 2.2 Disks in two dimensions

In this subsection Eqs. (4) and (6) are compared with simulation results of binary disks packed in two dimensions (*i.e.* the Euclidean plane). A Monte Carlo-based compression program was employed by Wan and Yang [34] to simulate the binary packing fraction. Monte Carlo methods have been extensively used in prior studies to investigate dense random packing structures. For instance, Chen et al. [35] generated truncated tetrahedra in maximally random jammed states using a sufficiently fast compression algorithm.

| $c_L$ | $\eta$ u = 1.4 | $\eta$ u = 1.7 | $\eta$ u = 2 | $\eta$ u = 3 |
|---|---|---|---|---|
| 0.3 | 0.84529 | 0.84801 | 0.85129 | - |
| 0.4 | 0.84532 | 0.84901 | 0.85285 | 0.86376 |
| 0.5 | 0.84586 | 0.85012 | 0.85412 | 0.86638 |
| 0.6 | 0.84590 | 0.85073 | 0.85520 | 0.86882 |
| 0.7 | 0.84563 | 0.85066 | 0.85504 | 0.86933 |
| 0.8 | 0.84522 | 0.84998 | 0.85324 | 0.86709 |
| 0.9 | - | - | 0.85104 | 0.85959 |

Table I Computationally generated binary packing fraction, $\eta(u, c_L, D = 2)$, of disks [34].

Meng et al. [36] produced dense random packings of monodisperse and binary spherocylinders by starting with configurations containing significant particle overlaps, followed by a relaxation algorithm. Wan and Yang [34] designed an algorithm based on fast compression that permits particle overlap, implemented using HOOMD-blue [37]. Specifically, they began with a random distribution of binary hard disks at a low packing fraction within a square box with periodic boundary conditions. A random compression factor between 0.9 and 1 was then selected, with which the box was compressed. If the resulting overlap, measured as the ratio of overlapping particles to the total particle count, was below 0.1, the compression was accepted, and overlaps were resolved using random Monte Carlo moves. Otherwise, the compression was rejected, and a new compression factor was chosen. This iterative process continued until dense packing configurations were achieved. In Table I the resulting packing fractions are included as function of size ratios u = 1.4, 1.7, 2 and 3 and of large disk volume.

| $X_L$ | $c_L$ u = 1.4 | $\eta$ u = 1.4 | $c_L$ u = 1.7 | $\eta$ u = 1.7 |
|---|---|---|---|---|
| 0.3 | 0.4565 | 0.84646 | 0.5533 | 0.85163 |
| 0.4 | 0.5665 | 0.84667 | 0.6583 | 0.85099 |
| 0.5 | 0.6622 | 0.84617 | 0.7429 | 0.84959 |
| 0.6 | 0.7462 | 0.84584 | 0.8126 | 0.84981 |
| 0.7 | 0.8206 | 0.84543 | 0.8709 | 0.84873 |

| $X_L$ | $c_L$ u = 2 | $\eta$ u = 2 |
|---|---|---|
| 0.3 | 0.6316 | 0.85417 |
| 0.4 | 0.7273 | 0.85411 |
| 0.5 | 0.8000 | 0.85340 |
| 0.6 | 0.8571 | 0.85196 |
| 0.7 | 0.9032 | 0.85026 |

Table II Computationally generated binary packing fraction, $\eta(u, c_L, D = 2)$, of disks [38].

Furthermore, binary disk packings are generated by Desmond [38], with the same algorithm reported in [39], but with a different energy minimizer. In Table II the generated packing fractions are tabulated for size ratios u = 1.4, 1.7 and 2, and for a number of number fractions $X_L$. With

$$c_L = \frac{X_L u^2}{1 + X_L(u^2 - 1)} \quad , \quad (7)$$

the large disk surface fraction is computed and is included in Table II as well. Note that $u^2$ is the surface area ratio of large and small disks.

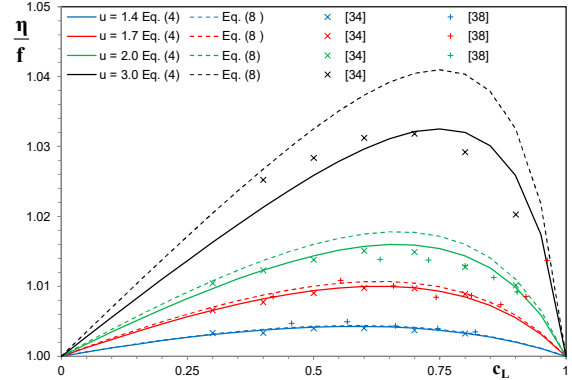

FIG. 1. Scaled packing fraction of randomly packed binary disks, $\eta(u, c_L, D = 2)/f$, versus large volume fraction $c_L$ and for size ratios u = 1.4, 1.7, 2 and 4, using model expressions Eq. (6) with either Eqs. (4) or (8), and the simulation values of Tables I [34] and II [38], using f = 0.8425.

The range of concentrations summarized in Tables I and II is such that the simulation protocols did not result in crystallization, which was found to be the case for lower and higher large disk fractions.
In Fig. 1, Eqs. (4) and (6), as well as the data of Tables I and II, are set out, scaled by a monosized packing fraction f of 0.8425. This value is based on fitting Eqs. (4) and (6) to the u = 1.4 data from both [34] and [38], which result in the same f, and this aforementioned f lies in the range of reported values [40], and is close to the value of 0.844 reported in [21]. In [40] the packing of monosized disks in a plane was studied employing a statistical geometric approach as well.

Fig. 1 shows that the data generated by [34] and [38] are compatible with each other, and that Eqs. (4) and (6) match very well with these packing simulations.

As said, the contraction function given by Eq. (4) follows from the combination of [10] and the excluded volume model [33] and this provided better predictions for larger u than the contraction function based on excluded volume only [32]. In Fig. 1, also Eq. (6) with as alternative contraction function [32]

$$v(u, D = 2) = (u - 1)^2 \quad , \quad (8)$$

is included. This function follows from the excluded volume model, which is correct near u = 1 [32]. In [32] it was shown that Eq. (4), which is an extension of the excluded volume solution to larger size ratios u, matches better with experimental data in $\mathbb{R}^3$ [10] than the original excluded volume expression. This extension converges to the excluded volume expression for u ↓ 1. Also in $\mathbb{R}^2$, Eq. (4) converges to Eq. (8) for D = 2 and u ↓ 1 [32] (Appendix).

Indeed for u close to unity, the use of Eqs. (4) and (8) leads to almost identical η/f, as expected, but for larger u, Eq. (8) tends to overestimate the binary packing fraction. The same trend was observed when applying the two different contraction functions to the packing of spheres in D = 3 [32]. The presented comparison of simulations and model confirms that the excluded volume approach of Onsager is applicable to disks in D = 2 and u up to 3 or so ($u^D ≈ 9$), and that extended contraction function Eq. (4) is most suitable indeed to capture the effect of size ratio on packing fraction for larger u. Also, the factor 1 – f in Eqs. (3) and (6), which followed from the excluded volume model, is a major factor in this equation. For the considered two-dimensional packing its value (≈ 0.16) is very distinct from the values pertaining to RLP (≈ 0.45) or RCP (≈ 0.36) of spheres packed in three dimensions [32]. As seen before in [32], the product of (1 – f) and Eq. (4) provides an accurate prediction of the packing increase by introducing bidispersity.

**2.3 Hyperspheres in infinitely large dimension**

Binary mixtures of hyperspheres with D → ∞ were studied by Ikeda *et al.* [30], constructing a statistical mechanical mean-field theory, based on the replica liquid theory to determine the fluid-glass transition in high-dimensions. Interestingly, the mean-field number density corresponds to the average number of overlaps counted in the excluded volume [41].

The monosized packing fraction f of these hyperspheres tends to zero in the large-dimension limit: f = $2^{1-D}$ (0.023 $D^2$ + 0.61 D + 0.365) [41], f ~ $2^{-D}$ (D log D) [42], f = $2^{1-D}$ (1.28 D - 1.36) [43] and ], f ~ $2^{-D}$ ($D^2$) [44], so

$$\lim_{D \to \infty} f = 0 \quad . \quad (9)$$

A scaling relation between size ratio u to the dimension D was introduced in [30] as follows

$$u = 1 + \frac{R}{D} \quad , \quad (10)$$

so that in the large-dimension limit holds

$$\lim_{D \to \infty} u^D = e^R \quad . \quad (11)$$

In Fig. 2, the scaled bimodal packing fraction, η(u, $c_L$, D)/f, following from Eqs. (4), (6), (9) and (11), is set out against the large hypersphere volume fraction $c_L$, employing R = 1/2.

| $c_L$ | η/f | $c_L$ | η/f |
|---|---|---|---|
| 0 | 1 | 0.5487 | 1.01573 |
| 0.0487 | 1.00231 | 0.5973 | 1.01564 |
| 0.1018 | 1.00453 | 0.6504 | 1.0152 |
| 0.1504 | 1.00667 | 0.6991 | 1.0144 |
| 0.1947 | 1.00836 | 0.7478 | 1.01324 |
| 0.2478 | 1.01022 | 0.7965 | 1.01173 |
| 0.3009 | 1.01173 | 0.8496 | 1.0096 |
| 0.3496 | 1.01307 | 0.8982 | 1.00684 |
| 0.3982 | 1.01413 | 0.9469 | 1.00391 |
| 0.4469 | 1.01484 | 1 | 1 |
| 0.5000 | 1.01547 | - | - |

Table III Scaled binary packing fraction, η($u^D = \sqrt{e}$, $c_L$, D → ∞)/f, of binary hyperspheres that followed from modelling, extracted from "Fig. 3" [30].

In this figure also the computational results from [30] are included, taken from "Fig. 3" (in which the scaled binary glass transition density is set out against small hypersphere volume fraction $c_S$, obviously $c_L + c_S = 1$), and which values are listed in Table III. In Fig. 2 an excellent agreement can be observed between the models presented here and in [30].

Ikeda *et al.* [30] provided data for R = 1/2, so $u^D = \sqrt{e}$ (≈ 1.649), which is smaller than 8 or 9, the limiting value for the model in D = 2 and 3. Hence, it appears that the results from statistical mechanical mean-field theory can also be explained with a geometric hard sphere packing model. Fig. 2 furthermore shows, as also found in [30], that though f is zero, the ratio η/f is not.

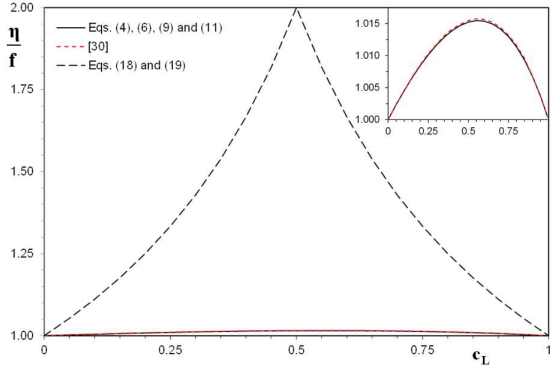

FIG. 2. Scaled packing fraction of binary hyperspheres, $\eta(u, c_L, D \rightarrow \infty)/f$, versus large volume fraction $c_L$, using model Eqs. (4), (6), (9) and (11), with R = 1/2, and model Eqs. (18) and (19) for R → ∞ (equal to "Eqs. (27) and (28)" [30]). The model results from "Fig. 3" of Ikeda *et al.* [30], listed in Table III, for R = 1/2 ($u^D = \sqrt{e}$) are also included. The inset shows a magnified view of the same graph for small $\eta(u^D = \sqrt{e}, c_L, D \rightarrow \infty)/f$.

### 3. LARGE SIZE RATIO

In this section the classic model of Furnas is revisited and applied to hyperspheres. This model provides closed-form expressions for binary particle mixes with infinite large size ratio (u → ∞).

#### 3.1 Furnas model

Furnas [4, 5] studied binary systems in three dimensions and it was concluded that the greater the difference in size between the two components, the greater the decrease in void volume. For infinitely large size ratio, the small particles fill the voids of the large particles, and they form separate and noninteracting phases. For this situation Furnas provided closed-form expressions.

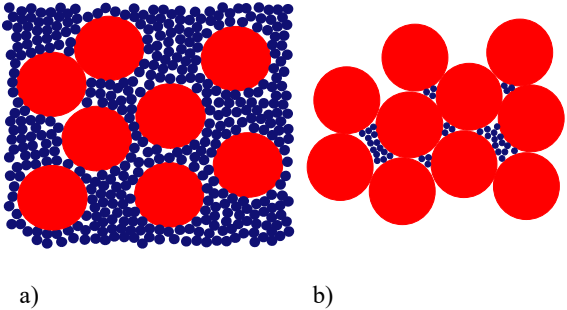

a)                   b)

FIG. 3 Binary packing of noninteracting particles (a) Larger spheres (or disks in D = 2) added to a monosized packing of smaller ones (b) Smaller spheres (or disks in D = 2) added to a monosized packing of larger ones.

The underlying concept also applies to combinations of two particle types that have different monosized packing fractions [45], *e.g.* because their shape is different, their particle size distributions differ, the mode of compaction differs, *etc.*. The concept is for instance also applicable to packings consisting of two continuously particle size distributions that are mixed [46]. It also applies to mixes of larger and smaller particles of which one or both of the phases are ordered (crystalline).

The only prerequisite is that the packing assembly of the smaller constituent fits in the open space between the larger one. Here, we will restrict ourselves to two monodisperse constituents that possess an identical packing fraction, which is the case for similar particles that are assembled identically.

The volume fraction of the large constituent is defined as

$$c_L = \frac{V_L}{V_L + V_S} \quad , \quad (12)$$

whereby for a binary packing fraction obviously holds

$$\eta(u, c_L, D) = \frac{V_L + V_S}{V_T} \quad , \quad (13)$$

and where $V_L$ and $V_S$ are the volumes of the large and small constituents in the packing, respectively, and with $V_T$ as total volume of the packing (entire space), including the voids.

First, a monosized packing of small particles only is considered ($c_L = 0$), in which large particles are introduced (Fig. 3a). This is the situation of a particle packing of small particles and their intermediate voids, total volume $V_S/f$, to which a volume $V_L$ of large particles is added. The binary packing fraction therefore reads as follows:

$$\eta(u \rightarrow \infty, c_L, D = 3) = \frac{V_L + V_S}{V_L + V_S/f} = \frac{f}{1 - c_L(1 - f)} \quad , \quad (14)$$

whereby Eq. (12) has been used.

Next, a packing of monosized large particles is considered, ($c_L = 1$), total volume $V_L/f$, to which small particles are added (Fig. 3b). The binary packing fraction reads as follows

$$\eta(u \rightarrow \infty, c_L, D = 3) = \frac{V_L + V_S}{V_L/f} = \frac{f}{c_L} \quad , \quad (15)$$

where again Eq. (12) has been used. Eqs. (14) and (15) intersect when large particles have the monosized packing fraction, and their voids are filled with small particles having the monosized packing fraction too. Furnas [4, 5] called mixes of binary particles that obey this composition "saturated mixtures", and in such mixtures sufficiently small particles are added to just fill the void fraction between the large particles. Large and small particles form two separate phases that have the same packing fractions, resulting the maximum binary packing fraction. For such saturated bidisperse packings, the volume fraction of the large fraction in the mix reads as follows:

$$c_{L,max} = c_L^{sat} = \frac{1}{2-f} \quad . \tag{16}$$

At this composition Eqs. (14) and (15) intersect, and

$$\frac{\eta_{max}}{f} = \frac{\eta^{sat}}{f} = 2 - f \quad , \tag{17}$$

whereby $\eta_{max}$ stands for $\eta(u \to \infty, c_L = c_{L,max}, D = 3)$, being the maximum for random binary packings or glasses. As this saturation point is the intersection of Eqs. (14) and (15), Eq. (14) is valid for $0 \leq c_L \leq c_L^{sat}$, and Eq. (15) for $c_L^{sat} \leq c_L \leq 1$. This saturation point can also be understood in another way: the large particles fill the total space with packing fraction f, and their voids are filled with the small particles' packing that subsequently occupy $(1 - f)f$ of the total space. Hence, both ingredients fill $(2 - f)f$ of the total space (Eq. (17)) and the volume fraction of large particles, $c_{L,max}$, is $(2 - f)^{-1}$ in this mix (Eq. (16)). Furthermore, mathematically, $(2 - f)f$ cannot exceed unity as $f \leq 1$.

Obviously, this concept is applicable only when the smaller ones do not affect the packing of the larger size group. Experiments with mixtures of broken solids [4, 5] and steel balls [11] revealed that noninteraction between subsequent size groups is obviously true when $u \to \infty$, but that non-disturbance is also closely approximated when $u \approx 7\text{-}10$. For angular particles, Caquot [9] found empirically a comparable size ratio ($u \approx 8\text{-}16$). Simulations showed that Eqs. (14) and (15) are approached closely for $u = 10$ [22, 25].

### 3.2 Hyperspheres in infinitely large dimension

The underlying concept, that the holes of the larger group are filled with the particles of the smaller groups, also holds for the D = 2 case (circles in a plane), to which Fig. 3 also applies. This spatial or geometric concept holds for all particle shapes, and for all modes of packing, from RLP to RCP. Hence, it stands to reason that this geometric concept also holds in higher dimensions, that is for hyperspheres (D > 3). This hypothesis is tested by an application to binary hyperspheres in the large-dimension limit, for which f = 0 [41-44], see Eq. (9).
Hence, it follows from Eqs. (14)-(16) that $c_L^{sat} = ½$, and that

$$\frac{\eta(u \to \infty, c_L)}{f} = \frac{1}{1 - c_L} \quad (0 \leq c_L \leq ½) \quad , \tag{18}$$

$$\frac{\eta(u \to \infty, c_L, D \to \infty)}{f} = \frac{1}{c_L} \quad (½ \leq c_L \leq 1) \quad . \tag{19}$$

The first equation was presented as "Eq. (28)" in [30]:

$$\frac{\eta(u \to \infty, c_L, D \to \infty)}{f} = \frac{1}{1 - c_L(1 - 2e^{-R/2})} \quad (0 \leq c_L \leq c_{L,max}) \quad , \tag{20}$$

when $R \to \infty$ is applied, and Eq. (19) corresponds to "Eqs. (27)" [30]. The maximum packing fraction is attained at $c_{L,max}$, which equals $c_L^{sat}$ (= ½) for $R \to \infty$. Eq. (20) also reveals that Eq. (14) is approximated with $e^{-R/2}$, so $u^{-D/2}$ [30], see Eq. (11). In the next section this limit will be explored in more detail, there a detailed study is presented of the asymptotic behavior of the binary packing fraction for large u.

The comparison with the results of [30] confirms the conjecture that the Furnas concept of noninteracting binary particles with large size ratio also holds for higher dimensions. Alternatively, one can say that the theoretical results by Ikeda *et al.* [30] for binary hyperspheres can be explained by the classic geometric concept of Furnas, originally developed for particles in three dimensions.

In Fig. 2, Eqs. (18) and (19) (or "Eqs. (27) and (28)" with applying $R \to \infty$ [30]) are set out. Again, though f is zero, the ratio $\eta/f$ is not, and in the large-dimension and large size ratio limits, its maximum $\eta^{sat}/f$ amounts to 2 at composition $c_L^{sat} = ½$ (Eqs. (16) and (17)). So where f scales with $2^{1-D}$ [41-44], $\eta^{sat}$ scales with $2^{2-D}$.

### 4. ASYMPTOTC APPROXIMTION FURNAS MODEL

Eqs. (3), (4) and (6) reveal that near u = 1 the binary packing varies with $(u^D - 1)^2$, this asymptotic behavior for u close to unity was discussed in detail in [32]. It is also interesting how the packing fraction approaches asymptotically the other limit, *viz.* $u \to \infty$ or $u^{-1} \downarrow 0$. Here, the Furnas model is extended by providing an asymptotic expansion of the binary packing fraction for $u^{-1}$ tending to zero.

### 4.1 Large particles added to small particles packing

First, the following normalized binary packing fraction is introduced

$$\lambda(u, c_L, D) = \frac{\eta(u, c_L, D) - f}{\eta(u, c_L, D)(1 - f)} \quad . \tag{21}$$

In [32] this transformation was introduced as for small size ratio $\lambda(u, c_L, D)$ does no longer depend on monosized packing fraction f. This readily follows from substituting Eq. (3) in Eq. (21). The RLP and RCP packing fractions of binary spheres with small size ratio indeed collapse when normalized by Eq. (21) [32]. These RLP and RCP assemblies have a very distinct factor $1 - f$ indeed (note that $1 - f$ is the void fraction of the monosized packing fraction).

In the previous section we have seen that in the large-dimension limit, in terms of $\lambda$ (Eq. (21)), Ikeda et al. [30] proposed the following approximation

$$\lambda(u, c_L, D \to \infty) = c_L(1 - \alpha u^{-\beta})$$

$$(0 \leq c_L \leq c_{L,\max}), \quad (22)$$

with $\alpha = 2$ and $\beta = D/2$, and which is based Eqs. (20) and (21). The large hypersphere volume fraction $c_{L,\max}$ is the volume fraction at which the binary packing fraction reaches the maximum. As seen in Section 3.1, $c_{L,\max}$ corresponds to $c_L^{sat}$ for $u \to \infty$, for which $(c_L^{sat}, \lambda^{sat}) = ((2-f)^{-1}, (2-f)^{-1})$. For $c_{L,\max} \leq c_L \leq 1$ no expansion of Eq. (19) in u was provided by [30].

| $c_L$ | $\eta$ u = 5 | $\lambda$ u = 5 | $\eta$ u = 10 | $\lambda$ u = 10 |
|---|---|---|---|---|
| 0 | 0.6435 | 0 | 0.6435 | 0 |
| 0.2 | 0.6761 | 0.135 | - | - |
| 0.4 | 0.7152 | 0.281 | 0.7278 | 0.325 |
| 0.5 | - | - | 0.7557 | 0.416 |
| 0.6 | 0.7525 | 0.406 | 0.7835 | 0.501 |
| 0.7 | 0.7714 | 0.465 | 0.8150 | 0.590 |
| 0.75 | - | - | 0.8270 | 0.622 |
| 0.8 | 0.7769 | 0.482 | 0.7948 | 0.190 |
| 0.9 | 0.7111 | 0.267 | - | - |
| 1 | 0.6435 | 0 | 0.6435 | 0 |

Table IV Binary packing fraction, $\eta(u, c_L, D = 3)$, and normalized binary packing fraction, $\lambda(u, c_L, D = 3)$, of binary spheres, with two different size ratios u, taken from "Table I" [22], that were computationally generated. The monosized packing fraction f is the value listed at $c_L = 0$ and $c_L = 1$ ($f = 0.6435$).

The governing variable in the large size difference limit is the magnitude of the power $\beta$. The infinite size ratio limit of Furnas is approached when $u \to \infty$, in that case the size of the voids that contain the small particles is infinitely larger than the small particle size, and the small particles attain their infinite volume monosized packing fraction. This void (or container) size scales linearly with the large particle size.

The relation between container size and monosized packing fraction was already studied by Scott [47] for monosized spheres. Scott [47] found that the packing fraction decreases with container size, and that the infinite packing fraction is approached by $u^{-1}$.

Desmond and Weeks [39] studied the effect of container size on monosized packing fraction, both for $D = 2$ (disks in $\mathbb{R}^2$) and $D = 3$ (spheres in $\mathbb{R}^3$), also yielding a $u^{-1}$ dependency in both dimensions. Based on analogical reasoning, it is invoked that this dependency holds for all dimensions, so $\beta = 1$. Moreover, for $u \to \infty$, in Section 3 we have seen that this analogical reasoning also held when applying the noninteraction model of Furnas (which corresponds to $u^{-1} = 0$) to $D \to \infty$.

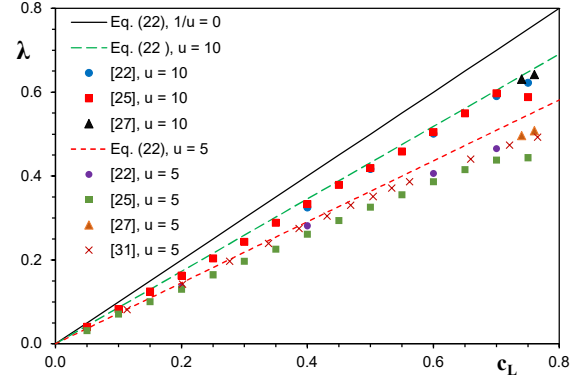

FIG. 4 Normalized binary random packing fraction $\lambda(u, c_L, D = 3)$ (defined by Eq. (21)) versus large volume fraction c as given by model Eq. (22) for u = 5, 10 and $\infty$, with $\alpha = 1.365$ and $\beta = 1$, and the computational values provided by [22, 25, 27, 31], listed in Tables IV, V, VI and VII. As $c_L^{sat} = (2-f)^{-1} \approx 0.733$, the tabled values $0 < c_L \leq 0.76$ are set out.

The asymptotic behavior asserted here, that is proportional to $u^{-1}$, is different from [30], which derived a $u^{-D/2}$ expansion. This latter expansion would imply that the Furnas limit can also be attained with a small size ratio u if D is large, which ignores the role of the size ratio.

The proposed asymptotic behavior toward the large size ratio limit is further analyzed by using the computational results of [22], [25], [27] and [31], concerning spheres in three dimensions. In Tables IV, V, VI and VII their binary RCP packing fraction results ($\eta$) for u = 5 and 10 are summarized, as well as the pertaining $\lambda$.

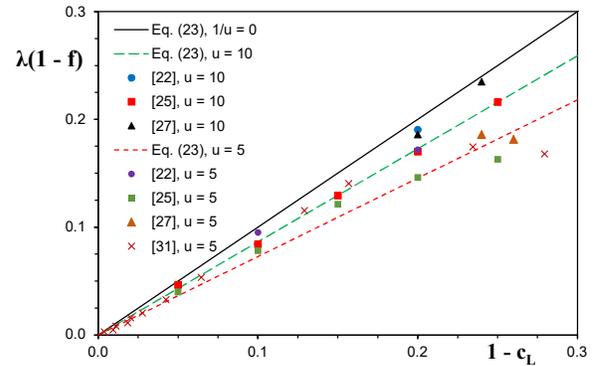

FIG. 5 Normalized binary random packing fraction $(1 - f)\lambda(u, c_L, D = 3)$ (defined by Eq. (21)) versus small volume fraction $c_S (= 1 - c_L)$ as given by model Eq. (23) for u = 5, 10 and $\infty$, with $\alpha = 1.365$, and the computational values provided by [22, 25, 27, 31], listed in Tables IV, V, VI and VII. As $c_L^{sat} = (2-f)^{-1} \approx 0.733$, the tabled values $0.72 \leq c_L < 1$ are set out.

In Fig. 4, Eq. (22) is set out for α = 1.365 (fitted) and β = 1, for u = 5, 10 and ∞, as well as the data from Tables IV, V, VI and VII concerning $0 \leq c_L \leq c_{L,max}$. One can see that Eq. (22) with α = 1.365 and β = 1 is able to capture the simulation results very well, especially for u = 10.

Hence, the asymptotic expansion proposed by Ikeda et al. [30], based on D → ∞, is also applicable to D = 3. The fitted value of α is such that it equals 2 − f considering that f ≈ 0.635 for RCP of spheres in D = 3. This relation between α and f is furthermore supported by α = 2 for D → ∞, see Eqs. (20) and (22), since then f = 0.

| $c_L$ | η u = 5 | λ u = 5 | η u = 10 | λ u = 10 |
|---|---|---|---|---|
| 0 | 0.633 | 0 | 0.633 | 0 |
| 0.05 | 0.640 | 0.032 | 0.642 | 0.040 |
| 0.1 | 0.650 | 0.071 | 0.653 | 0.083 |
| 0.15 | 0.657 | 0.101 | 0.663 | 0.124 |
| 0.2 | 0.665 | 0.130 | 0.673 | 0.163 |
| 0.25 | 0.673 | 0.165 | 0.684 | 0.204 |
| 0.3 | 0.682 | 0.196 | 0.695 | 0.243 |
| 0.35 | 0.690 | 0.225 | 0.708 | 0.288 |
| 0.4 | 0.700 | 0.261 | 0.721 | 0.334 |
| 0.45 | 0.709 | 0.294 | 0.735 | 0.379 |
| 0.5 | 0.719 | 0.325 | 0.748 | 0.419 |
| 0.55 | 0.728 | 0.355 | 0.761 | 0.458 |
| 0.6 | 0.738 | 0.387 | 0.777 | 0.505 |
| 0.65 | 0.746 | 0.415 | 0.793 | 0.549 |
| 0.7 | 0.754 | 0.438 | 0.810 | 0.597 |
| 0.75 | 0.756 | 0.163 | 0.807 | 0.216 |
| 0.8 | 0.741 | 0.146 | 0.762 | 0.170 |
| 0.85 | 0.720 | 0.121 | 0.727 | 0.129 |
| 0.9 | 0.686 | 0.078 | 0.691 | 0.084 |
| 0.95 | 0.659 | 0.040 | 0.664 | 0.046 |
| 1 | 0.633 | 0 | 0.633 | 0 |

Table V Binary packing fraction, η(u, $c_L$, D = 3), and normalized binary packing fraction, λ(u, $c_L$, D = 3), of binary spheres, with two different size ratios u, extracted from "Fig. 2" [25], that were computationally generated. The monosized packing fraction f is the value listed at $c_L$ = 0 and $c_L$ = 1 (f = 0.633).

| $c_L$ | η u = 5 | λ u = 5 | η u = 10 | λ u = 10 |
|---|---|---|---|---|
| 0.74 | 0.775 | 0.496 | 0.824 | 0.824 |
| 0.76 | 0.779 | 0.508 | 0.829 | 0.829 |
| 0.80 | - | - | 0.779 | 0.160 |

Table VI Binary packing fraction, η(u, $c_L$, D = 3), and normalized binary packing fraction, λ(u, $c_L$, D = 3), of binary spheres, with two different size ratios u, extracted from "Fig. 6" [27], that were computationally generated. The monosized packing fraction f = 0.634 [27].

### 4.2 Small particles added to large particles packing

The computational results of [22], [25], [27] and [31] also enable an analysis of the asymptotic behavior for $c_{L,max} \leq c_L \leq 1$. In Fig. 5, λ(u, $c_L$, D = 3) (1 - f) instead of λ(u, $c_L$, D = 3) is set out for $c_{L,max} \leq c_L \leq 1$, the values again taken from Tables IV, IV and VI. For this latter range, λ(1 - f) is set out, as then the different monosized packing fractions, viz. 0.6435 [22], 0.633 [25], 0.634 [27] and 0.645 [31], can be accounted for.

Hence, in terms of λ, the approximation of Eq. (15) (or Eq. (19)) is written as

$$\lambda(u, c_L, D = 3) (1 - f) = (1 - c_L) (1 - \alpha u^{-1})$$

$$(c_{L,max} \leq c_L \leq 1) \quad , \quad (23)$$

so asserting a similar asymptotic behavior as in the concentration range $0 \leq c_L \leq c_{L,max}$.

| $c_L$ | η | λ | $c_L$ | η | λ |
|---|---|---|---|---|---|
| 0 | 0.645 | 0.000 | 0.765 | 0.782 | 0.492 |
| 0.114 | 0.665 | 0.082 | 0.843 | 0.751 | 0.397 |
| 0.202 | 0.680 | 0.142 | 0.871 | 0.730 | 0.326 |
| 0.276 | 0.694 | 0.196 | 0.935 | 0.682 | 0.151 |
| 0.338 | 0.705 | 0.239 | 0.958 | 0.668 | 0.093 |
| 0.386 | 0.715 | 0.274 | 0.972 | 0.659 | 0.057 |
| 0.431 | 0.724 | 0.304 | 0.980 | 0.656 | 0.044 |
| 0.469 | 0.731 | 0.331 | 0.982 | 0.653 | 0.031 |
| 0.504 | 0.737 | 0.351 | 0.989 | 0.651 | 0.023 |
| 0.534 | 0.743 | 0.371 | 0.991 | 0.648 | 0.013 |
| 0.562 | 0.748 | 0.386 | 0.996 | 0.647 | 0.008 |
| 0.659 | 0.765 | 0.440 | 1 | 0.646 | 0 |
| 0.721 | 0.776 | 0.473 | | | |

Table VII Binary packing fraction, η(u = 5, $c_L$, D = 3), and normalized binary packing fraction, λ(u, $c_L$, D = 3), of binary spheres, extracted from "Fig. 7" [31], that were computationally generated. The monosized packing fraction f is the value listed at $c_L$ = 0 and $c_L$ = 1 (f = 0.645).

In Fig. 5, Eq. (23) is displayed, again using α = 1.365, and the data from Tables IV, V, VI and VII concerning $c_{L,max} \leq c_L \leq 1$, so invoking the same values for α and β as in the range $0 \leq c_L \leq c_{L,max}$. Though the number of data points is less and more dispersed, we can see that Eq. (23) is able to capture the asymptotic behavior in this concentration range for u ≥ 10 well, and that again the value of α = 1.365 provides very good agreement.

### 4.3 Expanded Furnas model

The previous analysis allows for new expressions for the Furnas limit as function of size ratio u. Based on the above analysis of λ, asymptotic approximations for the binary packing fraction η of spheres RCP for small $u^{-1}$ are obtained by transforming normalized packing fraction λ back to the binary packing fraction η, using the inverse of Eq. (21):

$$\eta(u, c_L, D) = \frac{f}{1 - \lambda(u, c_L, D)(1 - f)} \quad .$$

(24)

Substituting Eqs. (22) and (23) yield

$$\eta(u, c_L, D) = \frac{f}{1 - c_L(1-f)(1-(2-f)u^{-1})} \quad (0 \leq c_L \leq c_{L,max}) \,, \quad (25)$$

and

$$\eta(u, c_L, D) = \frac{f}{1 - (1-c_L)(1-(2-f)u^{-1})} \quad (c_{L,max} \leq c_L \leq 1) \,, \quad (26)$$

applicable from u = 10 to ∞, and perhaps even from u = 5 to ∞, whereby f = 0.635. One can recognize the similarity between Eq. (20) (or "Eq. (28)" of [30]) and Eq. (25) when f = 0 (which is the case for D → ∞) and that here the expansion follows $u^{-1}$ instead of $u^{-D/2}$ ($e^{-R/2}$), as discussed before.

Obviously Eqs. (25) and (26) tend to the original equations of Furnas, Eqs. (14) and (15), respectively, for $u^{-1} \downarrow 0$, so u → ∞. Eqs. (25) and (26) allow for an assessment of the bidisperse packing fraction for small $u^{-1}$, from 0 up to 0.1 or so. In other words, they are applicable to large but finite values of u (i.e. u > 10).

For D = 3, it was seen that α = 1.365 and β = 1 are appropriate values for RCP of spheres. The expressions presented here might also be applicable for other particle shapes and other dimensions D if α would equal 2 – f indeed. This conjecture is supported by the large-dimension limit findings, for which α = 2 and f = 0.

## 5. BINARY RANDOM PACKING DIAGRAM

In the previous sections it was seen that the binary random packing fraction can be described with the same equations for all D, viz. Eqs. (4) and (6) for $u^D \downarrow 1$, and Eqs. (25) and (26) for u → ∞. In this section a general binary packing fraction graph in $\mathbb{R}^D$ is presented, using the suitable normalized binary random packing fraction λ (Eq. (21)).

### 5.1 Phase boundaries

By using Eq. (21), the binary packing fraction for $u^D \downarrow 1$, Eq. (6), can be transformed into $\lambda(u^D, c_L, D)$ as follows

$$\lambda(u, c_L, D) = \frac{c_L(1-c_L)v(u, D)}{c_L(1-u^D) + u^D} \,. \quad (27)$$

This normalized binary packing fraction no longer contains the monosized packing fraction f, as was observed in [32] where this transformed binary packing fraction was introduced.

In Fig. 6 this normalized packing fraction is included for $u^D = 8$ and using Eq. (4), e.g. the case of packed disks in $\mathbb{R}^2$ for which u = 2√2, or spheres in $\mathbb{R}^3$ for which u = 2, the latter case being analyzed in [32].

In contrast to for u → ∞, for $u^D \downarrow 1$, λ does not depend on the monosized packing fraction f, it is governed by $u^D$ and composition $c_L$ only. However, toward large u, the packing fraction depends on monosized packing fraction f (which in turn depends on D) and composition $c_L$ only, see Eqs. (22) and (23).

Fig. 6 therefore is a simplified graph of the full range of all possible amorphous binary particle packing fractions, applicable to all $\mathbb{R}^D$, particle types and densification (from RLP to RCP), as function of composition and size ratio.

For u → ∞, the upper boundary lines result from Eqs. (22) and (23) and these lines form a triangle. The top of the triangle is termed $\lambda^{sat}$, which follows from Eqs. (17) and (21) as

$$\lambda_{max} = \lambda^{sat} = \frac{1}{2-f} \,. \quad (28)$$

The coordinates of this top are therefore $(c_L^{sat}, \lambda^{sat}) = ((2-f)^{-1}, (2-f)^{-1})$. Remarkably, the normalized packing fraction is thus bound by two lines defined by 2 – f. The shaded area covers the range of possible random binary packing fractions, which depend on composition $c_L$, size ratio u and monosized packing fraction f only. The maximum achievable normalized packing fraction, $(c_L^{sat}, \lambda^{sat}) = ((2-f)^{-1}, (2-f)^{-1}) \approx (0.733, 0.733)$, is indicated.

As observed before, for f = 0 (which is the case for D → ∞) and $u^{-1} = 0$ (u → ∞), Eqs. (22) and (23) form an isosceles triangle, with $(c_L^{sat}, \lambda^{sat}) = (½, ½)$ as maximum of the normalized packing fraction. The height from horizontal base to apex is thus ½. In the $\eta(u, c_L, D)$ graph, the upper boundary is then formed by two convex curves (Fig. 2), see Eqs. (18) and (19), that are reflectionally symmetrical with respect to vertical line $c_L = ½$.

### 5.2 Extrema

Toward u ↓ 1, the extremum of Eq. (27) follows from the partial derivative of $\lambda(u, c_L, D)$, with respect to $c_L$:

$$\lambda_{c_L}(u, c_L, D) = \frac{(u^D(1-c_L)^2 - c_L^2)v(u, D)}{(c_L(1-u^D) + u^D)^2} \,. \quad (29)$$

Equating Eq. (29) to zero yields the large particle volume fraction $c_{L,max}$ which results in the maximum packing for a given size ratio u and dimension D:

$$c_{L,max} = \frac{u^{D/2}}{u^{D/2} + 1} \,. \quad (30)$$

For $u^D \downarrow 1$, this maximum packing fraction occurs at $c_{L,max} = ½$. It also follows that $c_{L,max}$ is larger than ½ for $u^D > 1$. This can be seen in Fig. 2 (line $u^D = e^R = \sqrt{e}$, inset) and Fig. 6 (line $u^D = 8$). Eq. (30) also follows from determining the maximum of Eq. (6), that is by differentiating with respect to $c_L$ and equating the derivative to zero. So, the maximum of $\lambda(u, c_L, D)$ and of $\eta(u, c_L, D)$ occur at the same $c_{L,max}$. The maximum binary packing fraction therefore follows from substituting Eq. (30) into Eq. (6), producing

$$\eta_{max}(u, c_{L,max}, D) = \frac{(1-u^D)(1+u^{D/2})+u^{D/2}(1+u^{D/2})^2}{(1-u^D)(1+u^{D/2})+u^{D/2}(1+u^{D/2})^2-(1-f)v(u,D)} , \quad (31)$$

and the maximum of the normalized binary packing fraction follows from substituting Eq. (30) in Eq. (27), yielding

$$\lambda_{max}(u, c_{L,max}, D) = \frac{v(u,D)}{(u^{D/2}+1)^2} . \quad (32)$$

For the other limit, *viz.* large size ratio u, the asymptotic approximations are such that the volume fraction at maximum packing, $c_{L,max}$, is $c_L^{sat}$ (Eq. (16)) for $u^{-1} < 0.1$ ($u > 10$, so not $u \to \infty$ only). This particular follows from the intersection of Eqs. (25) and (26). At this $c_{L,max}$, the maximum packing fraction amounts to

$$\eta_{max}(u, c_{L,max}, D) = \frac{f(2-f)}{1+(1-f)(2-f)u^{-1}} , \quad (33)$$

which is the packing fraction at the intersection of Eqs. (25) and (26).

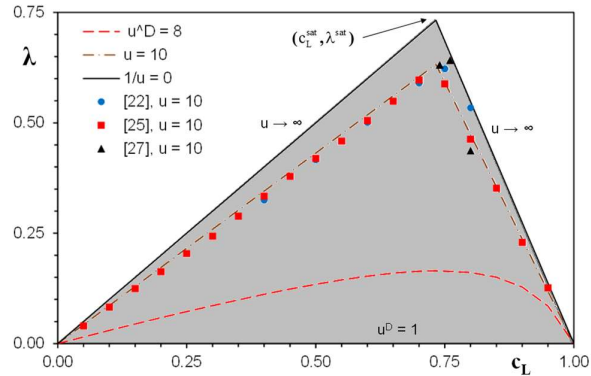

FIG. 6 Normalized binary random packing fraction $\lambda(u, c_L, D)$ (defined by Eq. (21)) for $u = 10$ and $u \to \infty$ (Eqs. (25) and (26)), for $u^D = 8$ (Eqs. (4) and (6)) and taking $f = 0.635$. The maximum binary packing fraction for $u \to \infty$, $(c_L^{sat}, \lambda^{sat}) = ((2 - f)^{-1}, (2 - f)^{-1}) \approx (0.733, 0.733)$, is indicated, which is also the maximum of all conceivable binary random packings.

At this intersection the maximum normalized packing fraction amounts to

$$\lambda_{max}(u, c_{L,max}, D) = \frac{1-(2-f)u^{-1}}{2-f} , \quad (34)$$

which the intersection of the lines given by Eqs. (22) and (23) at $c_{L,max} = c_L^{sat}$ (when $\alpha$ is taken to be $2 - f$ and $\beta = 1$).

Note that these $\eta_{max}$ and $\lambda_{max}$ do not depend on dimension D. For $u \to \infty$, $\lambda_{max}$ is $\lambda^{sat}$, and as concluded before, it appears that the values of $\lambda^{sat}$ and $c_L^{sat}$ are then the same, namely $(2 - f)^{-1}$, see Eq. (16) and Fig. 6. In Fig. 6, Eqs. (22) and (23) are set out with $\alpha = 2 - f$ and $\beta = 1$, and taking $f = 0.635$, for $u = 10$ and $u = \infty$, and the maxima can be seen.

Interestingly, for $f = 0$, $c_{L,max}$ is ½ for $u^D = 1$, then increases for $u^D > 1$, and for larger $u^D$ it returns to ½ again for u larger than 10 or so. Also for $f > 0$ one can observe that $c_{L,max}$ first increases with u increasing from unity ([11], [22] and [25]), and then decreases to $c_L^{sat}$ when u exceeds 10. For these cases this trend in $c_{L,max}$ eccentricity is more difficult to recognize than for $f = 0$, *i.e.* in the large-dimension limit.

## 6. CONCLUSION

This paper addresses the effect of bidispersity on the random packing fraction (or the glass transition), $\eta(u, c_L, D)$, of similar particles, with small size difference (u near unity) and large size difference ($u^{-1}$ near zero), in a variety of Euclidean spaces (D = 2, 3 and $\infty$), with emphasis on hyperspheres.

First, small size difference is studied. The model of [32], derived by combining Onsager's excluded volume model of particle pairs [33], and their statistical occurrence, and its validity extension by invoking the work of [10], is recapitulated. This modified excluded volume based model, governed by Eqs. (3) and (4), is successfully applied to both binary disks in $\mathbb{R}^2$, and to hyperspheres in the large-dimension limit. The approach is validated using the results of the theoretical [30] and computational studies [34, 38]. This model contains the factor $1 - f$, with f as monosized packing fraction. From D = 2 to D $\to \infty$ this factor varies from about 0.1575 to 1.

Subsequently, the opposite limit, infinite large size difference, is addressed. Here, the classic geometric model of Furnas is recapitulated. Furnas [4, 5] studied bidisperse mixtures of particle groups with constituents that have an infinitely large size disparity u. In this system the small and large particles form separate and noninteracting phases, the assembly of the small particles filling the voids of the assembly of the large particles. It is successfully demonstrated that this concept is also applicable to binary hyperspheres in the large-dimension limit, again using the theoretical results of [30].

An original model is proposed for the asymptotic approximation of the Furnas limit, so for $u^{-1} \downarrow 0$. It is reasoned that in all dimensions this approximation depends on $u^{-1}$. Based on the packing fraction of spheres in D = 3 and hyperspheres in D → ∞, there is sufficient evidence that the coefficient in the asymptotic approximation is 2 – f. New expressions for the binary packing fraction for large size ratio u are put forward, Eqs. (25) and (26), that converge to the original Furnas expressions, *viz.* Eqs. (14) and (15), for u → ∞, and which do not depend on space dimension D.

Finally, a packing graph of the normalized binary packing fraction $\lambda(u, c_L, D)$ is constructed (Fig. 6), featuring the limits of the amorphous state of D-dimensional binary packings as a function of a reduced number of parameters. This transformation of the packing fraction was introduced in [32] and does, in contrast to the binary packing fraction $\eta(u, c_L, D)$, not depend on monosized packing fraction f anymore for $u^D$ close to unity. The horizontal base line of the normalized packing fraction, to $u^D = 1$, is asymptotically approached by $(u^D – 1)^2$ with $u^D$ as volume ratio (surface ratio in $\mathbb{R}^2$) of the bidisperse particles.

Furthermore, for $u^{-1}$ close to zero, $\lambda(u, c_L, D)$ depends linearly on $c_L$, implying that the original packing fraction $\eta(u, c_L, D)$ are convex functions. In this limit, the normalized packing fraction depends on f, but not on space dimension D. The upper boundary of the normalized packing fraction, belonging to $u^{-1} = 0$, are formed by two straight lines, that are approached asymptotically by the term $(2 – f)u^{-1}$.

The $\lambda(u, c_L, D)$ figure features for u → ∞ the saturated composition, $c_L = c_L^{sat}$, which constitutes a special mix where the concentrations of large and small fractions are such that the small particles packing fits in the voids of the large particles packing (Fig. 3) and hence they form separate phases that have the same packing fraction. This combination of size ratio and composition $c_L$ also represents the maximum attainable random packing fraction of binary particles. It appears that the values of $\lambda^{sat}$ and $c_L^{sat}$ are the same, namely $(2 – f)^{-1}$, and these values characterize the graph (*e.g.* Fig. 6).

The packing fraction of the two monodisperse components, and of their bidisperse mix, depends on the compaction, which are asserted all to be identical. This compaction may correspond to MRJ, RCP, RLP, glass (transition) density *etc.*. The present model for the binary packing fraction can cope with all states of densification, *i.e.* they all feature the same binary packing fraction divided by monosized packing ($\eta(u, c_L, D)/f$). Though in the large-dimension limit the monodisperse packing fraction f of the hyperspheres tends to zero, $\eta(u, c_L, D)/f$ attains a finite value.

Hence, it appears that the packing fraction of the binary mix depends on the monosized packing fraction f, space dimension D, composition $c_L$ and size ratio u. In the small size disparity limit it depends on volume ratio $u^D$, and toward infinite size difference it depends on inverse size ratio $u^{-1}$.

Noteworthy, the models used are based on physical principles, and no adjustable parameter needed to be introduced to achieve the presented results. Concluding, one can say that the results of this study are a strong support for the applicability and validity of analytical "simple" and "athermal" hard hypersphere packing models, based on statistical geometry, to describe the amorphous states of soft condensed matter.

## APPENDIX: CONTRACTION FUNCTION

In this Appendix the contraction function v(u, D) as function of the space dimension is analyzed. Based on the application of the Onsager excluded volume concept to binary packings in arbitrary dimension D, the following contraction function was derived [32]

$$v(u, D) = w(u, D) (u - 1)^2 \quad , \tag{A.1}$$

with

$$w(u, D) = \frac{u^D + 1 - 2^{1-D}(u + 1)^D}{(1 - 2^{1-D})(u - 1)^D} \quad . \tag{A.2}$$

For D = 2, 3, …, 10, this equation was solved analytically and closed-form expressions for w(u, D) obtained [32]. These w(u, D) are polynomial functions in u of order D – 2, with all terms having positive coefficients. So each polynomial term $\alpha u^\beta$ ($\alpha > 0$, $2 \le \beta \le D – 2$) can expanded as $\alpha(1 + \beta(u – 1) + O((u – 1)^2)$ following the binomial series

$$(1+\varepsilon)^\beta = 1 + \beta\varepsilon + \frac{\beta(\beta-1)\varepsilon^2}{2!} + O(\varepsilon^3) \quad , \tag{A.3}$$

with $\varepsilon = u – 1$. This implies that near $\varepsilon = 0$ (in the vicinity of u = 1), the second order approximation of w(u, D) in (u – 1), see Eq, (A.1), equals w(u = 1, D) $(u – 1)^2$.

Eq. (A.1) is based on the Onsager excluded volume model, which is applicable to u close to unity. It was seen in [32] that for larger u (u values of around 2 to 3 or so) in $\mathbb{R}^3$ the following expression appeared to be more accurate

$$v(u, D) = w(u = 1, D) \frac{2(u^D - 1)^2}{D^2(u^D + 1)} \quad , \tag{A.4}$$

which follows from the experimental work by Mangelsdorf and Washington [10]. In Section 2.2 the same conclusion was drawn for $\mathbb{R}^2$.

It can be verified that Eqs. (A.1) and (A.4) converge for u ↓ 1 by using Eq. (A.3) and the asymptotic approximations

$$((1 + \varepsilon)^D - 1)^2 = (\varepsilon D)^2 + O(\varepsilon^3) \quad , \tag{A.5}$$

and

$$(1 + \varepsilon)^D + 1 = 2 + \varepsilon D + O(\varepsilon^2) \quad , \tag{A.6}$$

In other words, Eq. (A.4) features an application extension of the original excluded volume-based model for larger size ratios.

As said, in [32], w(u, D) was obtained for whole number dimensions D = 2, 3, …, 10, by solving Eq. (A.2). Subsequently, w(u = 1, D) was computed, which is needed to apply Eq. (A.4). It for instance followed that for D = 2 and D = 3, w(u = 1, D)/D² amounts to 1/4 and 2/9, respectively [32]. Also for D = 4, 5, …, 10, w(u = 1, D)/D² appeared to be close to 2/9.

| D | w(u = 1, D) | w(u = 1, D)/D² |
|---|---|---|
| 2 | 1 | 1.125 |
| 3 | 2 | 1 |
| 4 | 24/7 | 0.964 |
| 5 | 16/3 | 0.960 |
| 6 | 240/31 | 0.968 |
| 7 | 32/3 | 0.980 |
| 8 | 1792/127 | 0.992 |
| 9 | 1536/85 | 1.004 |
| 10 | 11520/511[+] | 1.014 |

Table A.1 Values of w(u = 1, D) following from Eq. (A.2), their scaled value, and the asymptotic approximation Eq. (A.9) for large D. [+]In [32] this value was erroneously listed as 11394/511.

In Table A.1 all computed values of w(u = 1, D) are listed, that follow from Eq. (A.2) [32], as well as the values scaled by D². In Fig. A.1 these scaled values for D = 2, 3, …, 10 are displayed.

As only w(u = 1, D) needs to be specified to apply Eq. (A.4), the limit of Eq. (A.2) for u ↓ 1 is determined:

$$w(u = 1, D) = \lim_{u \to 1} \frac{u^D + 1 - 2^{1-D}(u+1)^D}{(1 - 2^{1-D})(u-1)^2} = \frac{D(D-1)}{4(1 - 2^{1-D})}, \tag{A.7}$$

whereby L'Hôpital's (or Bernoulli's) rule is applied twice. This equation reveals that for D → ∞, w(u = 1, D)/D² tends to 1/4, the same value as for D = 2. In Fig. A.1 w(u = 1, D)/D², using Eq. (A.7), is also included.

It indeed follows that Eq. (A.7) indeed coincides with w(u = 1, D) that followed from solving Eq. (A.2) for D = 2, 3,.., 10.

Eq. (A.7) however provides an analytical expression for w(u = 1, D)/D² for all D, which is a continuous function. The $D_{min}$ pertaining to this minimum can be computed by differentiating w(u = 1, D)/D² by D, and equating the result to zero:

$$1 - 2^{1-D_{min}} + \text{Ln}(2)D_{min}(D_{min} - 1)2^{1-D_{min}} = 0 \quad . \tag{A.8}$$

Solving this implicit algebraic equation in $D_{min}$ yields the non-Euclidean space $D_{min} \approx 4.72$ for which w(u = 1, D = $D_{min}$)/$D_{min}^2 \approx 0.213$.

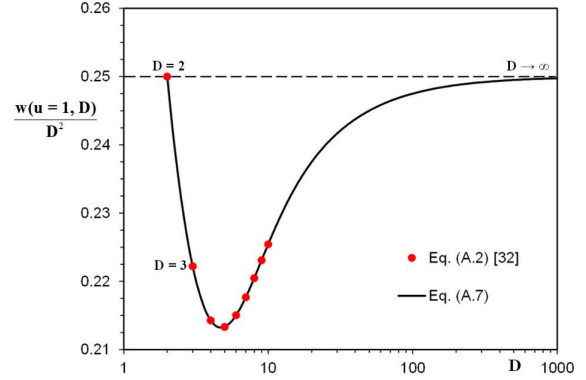

FIG. A.1 Scaled values of w(u = 1, D), w(u = 1, D)/D², as function of space dimension D. The solutions of Eq. (A.2) are taken from [32] and are also listed in Table A.1, the analytic solution (Eq. (A.7)) is also displayed. The horizontal asymptote w(u = 1, D)/D² = 1/4 is included to guide the eye.

To summarize, Table A.1 and Figure A.1 thus reveal that w(u = 1, D)/D² is maximum at D = 2, then decreases to $D_{min} \approx 4.72$, and then increases again toward the same maximum as D = 2, namely w(u = 1, D)/D² = 1/4. This asymptote is included in Fig. A.1.

In [32], w(u = 1, D)/D² was solved for integer values D = 2 up to D = 10 (Table A.1). As for D = 3 to 10 the values were very similar and close to 2/9 (see Fig. A.1), it was extrapolated that this w(u = 1, D)/D² = 2/9 also holds for D > 10. However, from the analysis presented here, we can conclude that for larger D, w(u = 1, D)/D² approaches 1/4 (Fig. A.1), that is the same value as for D = 2.

## ACKNOWLEDGMENTS

Dr. D.D. Wan and Mr. Y.H. Yang (Wuhan University), and Dr. K.W. Desmond (Emory University), are acknowledged for providing computer simulations of binary disk packings (Tables I and II, respectively). Dr. F. Zamponi (Sapienza University) is thanked for the discussions and providing Ref. [30]. Dr. Y. Luo is acknowledged for extracting the data of Refs. [25], [27], [30] and [31], displayed in Figs. 2, 4, 5 and 6. Mrs. Y.L. Chen is thanked for assisting in the drawing Fig. 3. The author dedicates this paper to the memory of his mother Elisabeth J.H. Brouwers (née Spauwen), 11 March 1921 (Eijsden) – 5 December 1999 (Maastricht), whose encouragement and sacrifices laid the foundation for this endeavor.

## DATA AVAILABILITY

The data that support the findings of this article are not publicly available. The data are available from the author upon reasonable request.

---

[1] S. Torquato and F.H. Stillinger, Rev. Mod. Phys. 82, 2633 (2010).
[2] G. Parisi and F. Zampini, Rev. Mod. Phys. 82, 789 (2010).
[3] P. Charbonneau, A. Ikeda, A., G. Parisi, G. and F. Zamponi, Phys. Rev. Lett. 107, 185702 (2011).
[4] C.C. Furnas, Department of Commerce, Bureau of Mines, Report of Investigation Serial No. 2894 (1928).
[5] C.C. Furnas; Bulletin of US Bureau of Mines 307, 74 (1929).
[6] A.E.R. Westman and H.R. Hugill, J. Am. Ceram. Soc. 13, 767 (1930).
[7] C.C. Furnas, Ind. Eng. Chem. 23, 1052 (1931).
[8] A.E.R. Westman, J. Am. Ceram. Soc. 19, 127 (1936).
[9] M.A. Caquot, Mémoires de la Société des Ingénieurs Civils de France, 562 (in French) (1937).
[10] P.C. Mangelsdorf and E.L. Washington, Nature, Lond. 187, 930 (1960).
[11] R.K. McGeary, J. Am. Ceram. Soc. 44, 513 (1961).
[12] R. Jeschar, W. Pötke, V. Petersen and K. Polthier, Proceedings Symposium on Blast Furnace Aerodynamics, Wollongong, Australia, 25-27 September 1975, Ed. N. Standish, Illawara Branch of the Australasian Institute of Mining and Metallurgy (Aus. I.M.M.) (1975), pp. 136-147.
[13] A.S. Clarke and J.D. Wiley, Phys. Rev. B 35, 7350 (1987).
[14] A.P. Shapiro and R.F. Probstein, Phys. Rev. Lett. 68, 1422 (1992).
[15] A. Yang, C.T. Miller and L.D. Turcoliver, Phys. Rev. E 53, 1516 (1996).
[16] D. He, N.N. Ekere and L. Cai, Phys. Rev. E 60, 7098 (1999).
[17] H.J.H. Brouwers, Phys. Rev. E 74, 031309, ibid 069901 (E) (2006), ibid E 84, 059905 (E) (2011).
[18] H.J.H. Brouwers, Phys. Rev. E 76, 041304 (2007).
[19] Y. Shi and Y. Zhang, Appl. Phys. A 92, 621 (2008).
[20] H.J.H. Brouwers, Phys. Rev. E 78, 011303 (2008).
[21] I. Biazzo, F. Caltagirone, G. Parisi and F. Zamponi, Phys. Rev. Lett. 102, 195701 (2009).
[22] R.S. Farr and R.D. Groot, J. Chem. Phys. 131, 244104 (2009).
[23] M. Clusel, E.I., Corwin, A.O.N. Siemens and J. Brujić, Nature 460, 611 (2009).
[24] M. Danisch, Y.L. Jin and H.A. Makse, Phys. Rev. E 81, 051303 (2010).
[25] A.V. Kyrylyuk, A. Wouterse and A.P. Philipse, Progr. Colloid Polym. Sci. 137, 29 (2010).
[26] K.A. Newhall, I. Jorjadze, E. Vanden-Eijnden and J. Brujić, Soft Matter 7, 11518 (2011).
[27] A.B. Hopkins, F.H. Stillinger, S. Torquato, Phys. Rev. E 88, 022205 (2013).
[28] H.J.H. Brouwers, Phys. Rev. E 87, 032202 (2013).
[29] K.W. Desmond and E.R. Weeks, Phys. Rev. E. 90, 022204 (2014).
[30] H. Ikeda, K. Miyazaki, H. Yoshino, A. Ikeda, Phys. Rev. E 103, 022613 (2021).
[31] C. Anzivino, M. Casiulis, T. Zhang, A.S. Moussa, S. Martiniani and A. Zaccone, J. Chem. Phys. 158, 044901 (2023).
[32] H.J.H. Brouwers, Physics – Uspekhi 67, 510 (2024).
[33] L. Onsager, Ann. N.Y. Acad. Sci. 51, 627 (1949).
[34] Y.H. Yang and D.D. Wan, Private communications (2025).
[35] D.Y Chen, Y. Jiao and S. Torquato, J. Phys, Chem. B 118, 7981 (2014).
[36] L. Meng, P. Lu, S. Li, J. Zhao, and T. Li, T. Powder Technol. 228, 284 (2012).
[37] J. A. Anderson, J. Glaser, and S. C. Glotzer. Comput. Mater. Sci 173, 109363 (2020).
[38] K.W. Desmond, Private communications (2025).
[39] K.W. Desmond and E.R. Weeks, Phys. Rev. E. 80, 051305 (2009).
[40] H.J.H. Brouwers, Soft Matter 19, 8465 (2023).
[41] A.P. Philipse, Colloids Surf. A 213, 167 (2003).
[42] G. Parisi and F. Zamponi, J. Stat. Mech.: Theory Exp., P03017 (2006).
[43] M. Skoge, A. Donev, F.H. Stillinger and S. Torquato, Phys. Rev. E 74, 041127 (2006).
[44] B. Schmid and R. Schilling, Phys. Rev. E 81, 041502 (2010).
[45] H.J.H. Brouwers, Phys. Rev. E 89, 052211 (2014).
[46] H.Y. Sohn and C. Moreland, Can. J. Chem. Eng. 46, 162 (1968).
[47] G.D. Scott, Nature, Lond. 188, 908 (1960).